# High pO$_2$ Floating Zone Single Crystal Growth of the Perovskite Nickelate PrNiO$_{3-x}$

Hong Zheng[1], Junjie Zhang[2], Bixia Wang[1], Daniel Phelan[1], M. J. Krogstad[1], Yang Ren[3], W. Adam Phelan[4], Omar Chmaissem[5], Bisham Poudel[5], J. F. Mitchell[1]

[1]Materials Science Division, Argonne National Laboratory, Argonne, Illinois 60439, USA

[2]Materials Science and Technology Division, Oak Ridge National laboratory, TN 37831, USA

[3]Advanced Photon Source, Argonne National Laboratory, Argonne, Illinois 60439, USA

[4]Platform for the Accelerated Realization, Analysis and Discovery of Interface Materials (PARADIM), Department of chemistry, The Johns Hopkins University, Baltimore, MD 21218, USA

[5]Department of Physics, Northern Illinois University, Dekalb, Illinois 60115

**Abstract:** Single crystals of PrNiO$_{3-x}$ were grown under an oxygen pressure of 295 bar using a unique high-pressure optical-image floating zone furnace. The crystals, with volume in excess of 1 mm$^3$, were characterized structurally using single crystal and powder X-ray diffraction. Resistivity, specific heat, and magnetic susceptibility. All of which evidenced an abrupt, first order metal-insulator transition (MIT) at ~130 K, in agreement with previous literature reports on polycrystalline specimens. Single crystal diffraction below 130 K revealed a monoclinic symmetry (space group $P2_1/n$), consistent with existing models of charge disproportionation as the driver for the MIT. Our study demonstrates the opportunity space for high fugacity, reactive environments for single crystal growth specifically of perovskite nickelates but more generally to correlated electron oxides.

**Introduction**

The perovskite nickelates, RNiO$_3$ (R=La, Pr-Lu), have received ongoing attention due to their complex magnetic, transport, and structural phase diagram [1-23]. Except for LaNiO$_3$ [1-3, 6, 24-30], all of the perovskite nickelates undergo a first-order metal-insulator transition (MIT) as a function of temperature. The transition temperature of the MIT depends on the size of the rare-earth ions, and correlates with the Ni-O-Ni bond angle [1-3]. The nature of the MIT has been discussed in the literature in terms of charge disproportionation [14, 16, 31-34], negative charge transfer [5], and bipolaron condensation [4], but no consensus has been reached. In addition to the MIT, antiferromagnetic order at $T_N$ occurs concomitantly with MIT for R= Pr$^{3+}$ and Nd$^{3+}$, and below $T_{MIT}$ for rare earth ions smaller than Nd. However, the magnetic structure of the ground state remains an open question. In particular, three magnetic structures have been proposed base on powder neutron diffraction that are consistent with an observed propagation vector q =(¼, ¼, ¼); however, no experiments have been able to differentiate among them [15]. Polycrystalline powders and thin films provide an incomplete means to understand the nature of the structural and magnetic transitions in the perovskite nickelates [1-2]. For instance, to test the potential magnetic structure candidates. Single crystals of size and quality suitable for neutron diffraction are necessary.

It is well-established that high oxygen partial pressure ($p$O$_2$) is needed to prepare polycrystalline specimens of RNiO$_3$ [35-37]. For single crystal growth, it was reported that single crystals of PrNiO$_3$ with dimensions of ~ 0.5 mm on an edge could be grown by slowly cooling the melt of Pr$_6$O$_{11}$ + 6NiO + 0.5 KClO$_4$ + 0.5 KCl + 0.5 NaClO$_4$ + 0.5 NaCl at 4.5 GPa by Saito et al. in 2003 [38-39]. Micrometer-sized single crystals of NdNiO$_3$ (~ 100 μm on an edge) were grown in a belt-type press at 4 GPa by Alonso et al. [40]. Recently, we reported the successful single crystal growth of LaNiO$_3$ using a high pressure floating zone furnace [27], and subsequently so did Guo et al. [26], reviving interest in this strongly correlated metallic oxide. The required pO$_2$ for stabilizing PrNiO$_3$ is expected exceed that of the LaNiO$_3$ case, as is true for polycrystalline synthesis of nickelates of the smaller rare-earth ions [36].

In this paper, we report successful crystal growth of PrNiO$_3$ at 295 bar of oxygen pressure via the floating-zone technique. Single crystal x-ray scattering demonstrates the quality of the single crystals and reveals temperature-induced changes near $T_{MI}$. Electrical resistance and heat capacity measurements on the grown single crystals evidence MITs and show how the MITs and the low-temperature states respond to oxygen stoichiometry.

**Experimental Section**

**Crystal Growth**. As-purchased $Pr_6O_{11}$ (Alfa Aesar, 99.99%) was baked at 600 °C for at least 48 h before use. A stoichiometric ratio of $Pr_6O_{11}$ and NiO (Alfa Aesar, 99.99%) powder was weighed, and the mixture was thoroughly ground, loaded into a Pt crucible, heated in flowing $O_2$ to 1050 °C at a rate of 3 °C/min, allowed to dwell for 24 h, and then furnace-cooled to room temperature. The solid was then reground and sintered twice at 1050 °C, with intermediate grinding using the same atmosphere and heating protocol. The powder was then hydrostatically pressed into polycrystalline rods (length ~ 80 mm, diameter ~ 6 mm) at 15,000 psi and sintered at 950 to 1050 °C for 24 h for crystal growth. $PrNiO_3$ single crystals were grown using the 300 bar high pressure floating zone furnace (SciDre GmbH) at the Platform for the Accelerated Realization, Analysis and Discovery of Interface Materials (PARADIM) user facility at the Johns Hopkins university. A 5 kW Xenon arc lamp was utilized as a heating source at pressure of 295 bar with an oxygen flow rate of 0.2 L/min. Feed and seed rods were counter-rotated at 20 and 15 rpm, respectively. The seed rod was advanced at a rate of 5 mm/h. Figure 2a shows crystal boules of $PrNiO_{3-x}$, and large single crystals can be seen in the polished cross section (see Figure 2b). The crystallinity of the of the as-grown boule was characterized using high-energy single crystal synchrotron X-ray diffraction at Beamline 11-ID-C ($\lambda$ = 0.11165 Å, beam size = 0.5 × 0.5 mm$^2$) at the Advanced Photon Source (APS) at Argonne National Laboratory, shown in Figure 2c.

**Powder X-ray Diffraction.** Powder X-ray diffraction data were collected at room temperature on pulverized single crystals using a PANalytical X'Pert Pro powder diffractometer with Cu K$_\alpha$ radiation ($\lambda$ =1.5418 Å) in the 2θ range of 5-80º. The data were analyzed using GSAS-II software using the structure model from the literature [35]. Refined parameters include scale factor, sample displacement, background (8$^{th}$ polynomial function), lattice parameters, isotropic domain size, isotropic microstrain, and isotropic thermal parameter $U_{iso}$ (constrained to be the same).

**Single Crystal X-ray Diffraction.** Single crystal X-ray diffraction was performed at Beamline 6-

ID-D at the Advanced Photon Source using a wavelength of 0.14238 Å and a 2M CdTe Pilatus detector. Data collection was performed under continuous rotation, and data were transformed from detector space to reciprocal space using NeXpy.

**Electrical Transport**. The resistivity of $PrNiO_3$ single crystals was measured on a Quantum Design PPMS in the temperature range 1.8-300 K using the van der Pauw method with Indium contacts.

**Specific Heat.** Specific heat measurements were performed on a Quantum Design PPMS in the temperature range 2 to 260 K. Apiezon-N vacuum grease was employed to fix ~20 mg crystals to the sapphire sample platform. Addenda were subtracted.

**Magnetic Susceptibility**. Magnetic susceptibility measurements were performed using a Quantum Design MPMS-3 SQUID magnetometer. A small piece of $PrNiO_3$ crystal cut from the crystal boule , was attached to a quartz rod using a minimum amount of Apiezon grease. Zero-field cooled (ZFC) and field cooled (FC) data were collected under a magnetic field of 0.1 T. The sample was cooled in zero field to 2 K at a rate of 35 K/min, after which the field was applied, and DC magnetization measurement was performed on warming at 2 K/min (ZFC). The sample was then cooled in field at 35 K/min to 2 K, and then DC magnetization data were collected on warming at 2 K/min (FC).

**Results and Discussion**

High oxygen fugacity environments are essential to prepare bulk $RNiO_3$ specimens [35-36], motivating attempts to grow single crystals under extreme high oxygen pressure. For instance, ~ 0.5 mm-size single crystals of $PrNiO_3$ were grown by by Saito et al. at 4.5 GPa [38], but the nature of the high pressure flux growth experiment put significant constraints on the obtained crystal size. Recently, we and others have found that $RNiO_3$ crystals can be grown at significantly lower $pO_2$ using a floating-zone approach at high oxygen fugacity. For example, $LaNiO_3$ crystals were grown by our group [27] and subsequently by the MPI-Dresden group [26] using high oxygen pressure ($pO_2$ ~ 30-150 bar). Unfortunately, our attempts to grow $PrNiO_3$ using the 150 bar floating zone furnace in our Laboratory were unsuccessful. However, crystal growth was successful using the 300 bar floating zone furnace installed at the NSF-sponsored PARADIM crystal growth facility located at Johns Hopkins University. Indeed, success in this case was serendipitous, as we observed the appearance of PrNiO$_3$ as a biproduct of growth of the trilayer Ruddlesden-Popper phase Pr$_4$Ni$_3$O$_{10}$ [41-42]. As shown in Figure 1, PrNiO$_3$ was observed as a minor phase at 220 bar pO$_2$, but as a significant component at 260 bar. The PrNiO$_3$ phase was not observed in growth at or below 180 bar. Thus, increasing pO$_2$ favors the perovskite phase (nominal Ni$^{3+}$) at the expense of the layered phase (nominal Ni$^{2.67+}$), consistent with our work on the La-case [41].

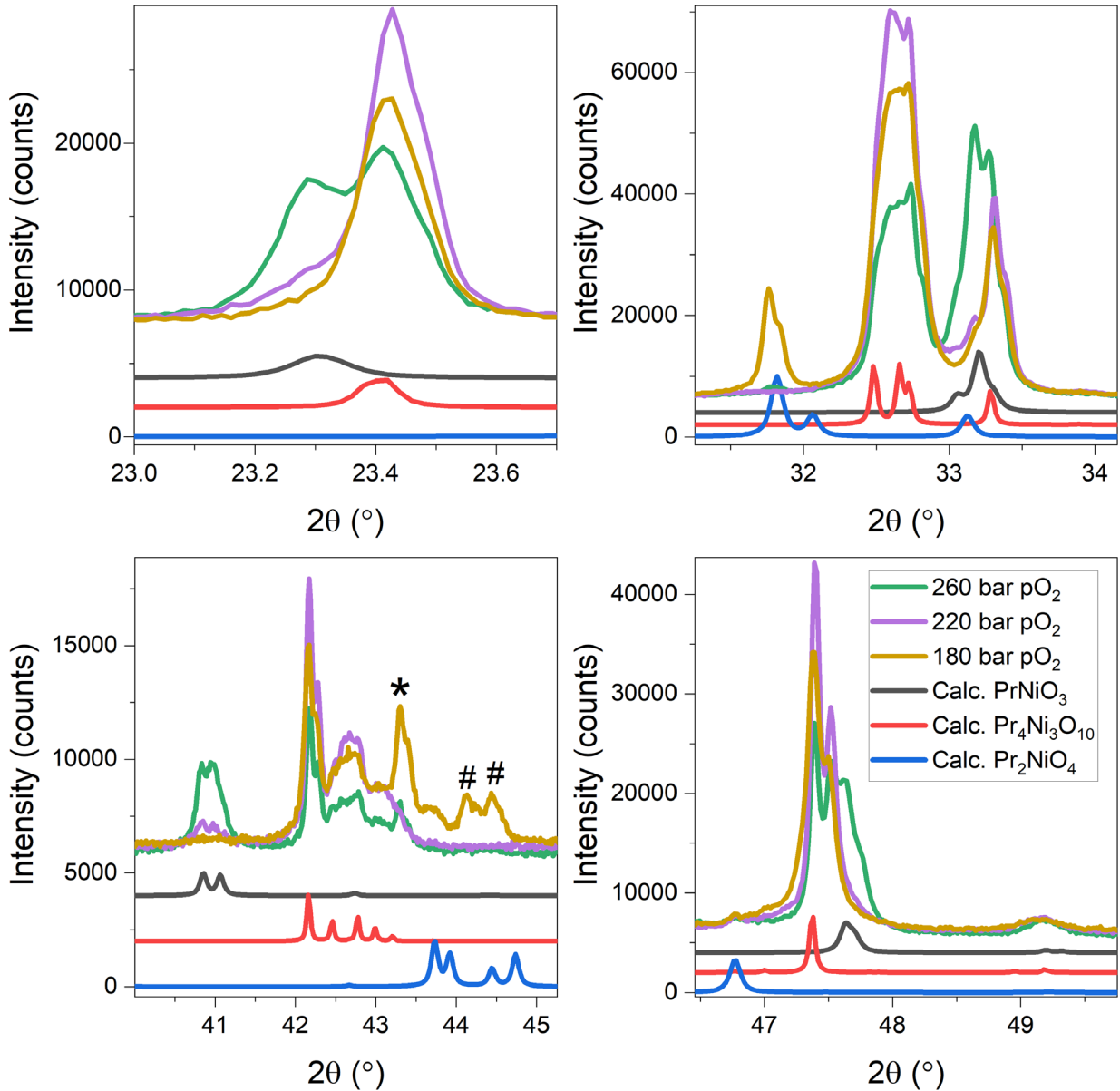

Figure 1. X-ray powder diffraction patterns of the phases obtained during floating zone growth under different pO$_2$ with a feed rod composition of Pr: Ni = 4:3. The PrNiO$_3$ phase is observed to increase in volume fraction upon increasing pO$_2$. Bottom three curves in each panel are calculated patterns for PrNiO$_3$, Pr$_4$Ni$_3$O$_{10}$, and Pr$_2$NiO$_4$, respectively. The * and # mark reflections from NiO and Pr$_6$O$_{11}$ second phases, respectively.

We then turned directly to the growth of PrNiO$_3$ with a molar ratio of Pr:Ni=1:1 at pO$_2$=295 bar of oxygen, which was close to the upper limit of the furnace (300 bar). The starting rod itself was not a single PrNiO$_3$ phase. Rather, it consisted of a mixture of PrO$_x$, NiO, and Pr$_2$NiO$_4$ because the PrNiO$_3$ phase is formed only at elevated oxygen pressure. During initial growth efforts, we found that stability of the zone was difficult to maintain both because of cracks that formed on the feed rod itself, prior to melting, and the low viscosity of the molten liquid in the zone, leading to frequent zone loss. To improve the stability and prevent the cracking in the feed rod, we found that preparing the rods at low sintering temperatures (950 °C) was beneficial, as severe cracks were less likely to appear when the feed rods were less dense. We surmise that small amounts of molten liquid penetrate into the feed rod and seal the cracks before they can expand catastrophically. Fortunately, excessive penetration of the liquid into the feed rod was not observed, which would have led to a deleterious 'pre-melt' zone often found with low-density feed rods.

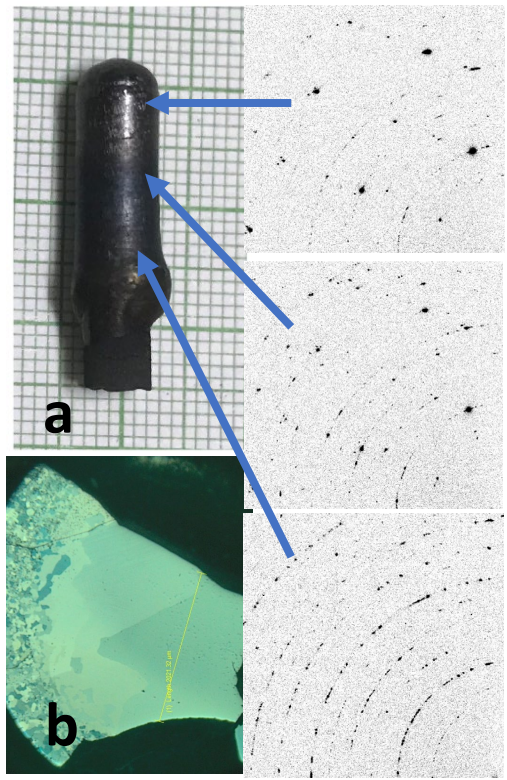

Figure 2. (a) PrNiO$_3$ crystal boule grown under 295 bar of oxygen. (b) A polished surface of a slice from the PrNiO$_3$ boule. (c) X-ray single crystal diffraction patterns measured at various locations along the length of the boule at 11-ID-C.

Figure 2a shows a PrNiO$_3$ crystal boule grown under 295 bar pO$_2$ at a traveling rate of 5 mm/h, and Figure 2b shows a part of polished cross section surface from the boule. It is apparent that the inside core of the boule (right side of Figure 2b) consists of several ~ mm size single crystals, whereas the outside of the boule consists of a polycrystalline shell. Despite several attempts, we have been unable to prepare single crystal boule. Both the single crystals on the inside core and the polycrystalline shell on the outside consist of the PrNiO$_3$ phase without any impurities, as adjudged by laboratory X-ray powder diffraction. The mechanism leading to this polycrystalline shell is not understood. It is not unreasonable to speculate that it emerges due to contact with the supercritical fluid growth environment [43], but verifying this will require further experiments. Figure 2c shows single crystal diffraction patterns measured in transmission mode along the length of the boule at the beamline 11-ID-C of the APS. The evolution from a ceramic rod to single crystals can be seen by the transition from a textured ring of scattering to well-defined Bragg reflections along the growth direction. The constellation of diffraction spots reflects the fact that the crystal boule consists of multiple grains, as shown in Figure 2b.

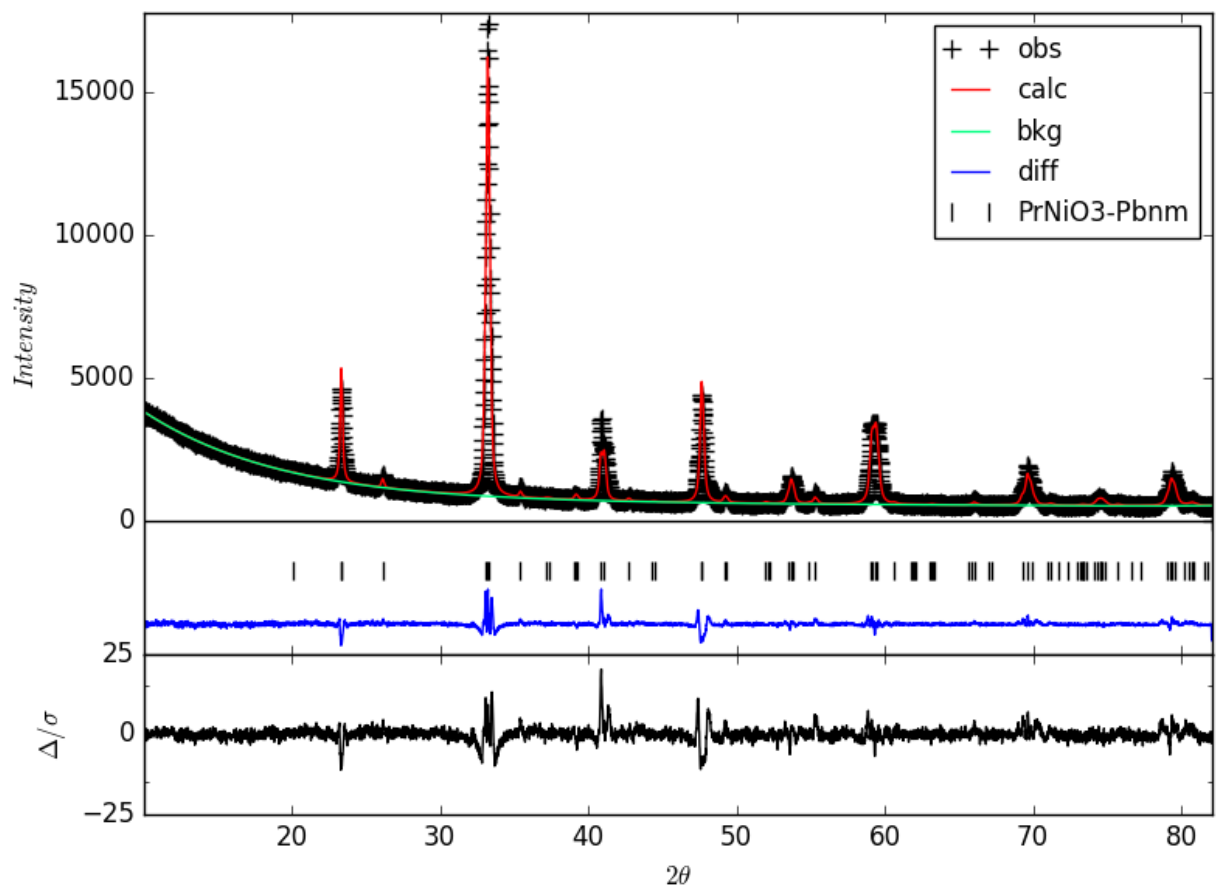

Figure 3. Rietveld refinement of laboratory-based x-ray powder diffraction of PrNiO$_3$ measured at room temperature.

Figure 3 shows Rietveld refinement of laboratory powder x-ray diffraction collected from a pulverized quarter disc cut from the crystal boule, representing a bulk average of the resultant growth. No second phases were found. All peaks can be indexed in the space group *Pbnm*, with lattice parameters a=5.4242 Å, b=5.3781 Å, c=7.6322 Å, which are comparable to the powder results reported by Lacorre et al. [35] for a polycrystalline specimen, a = 5. 4154 Å, b=5.3755 Å, c=7.6192 Å .

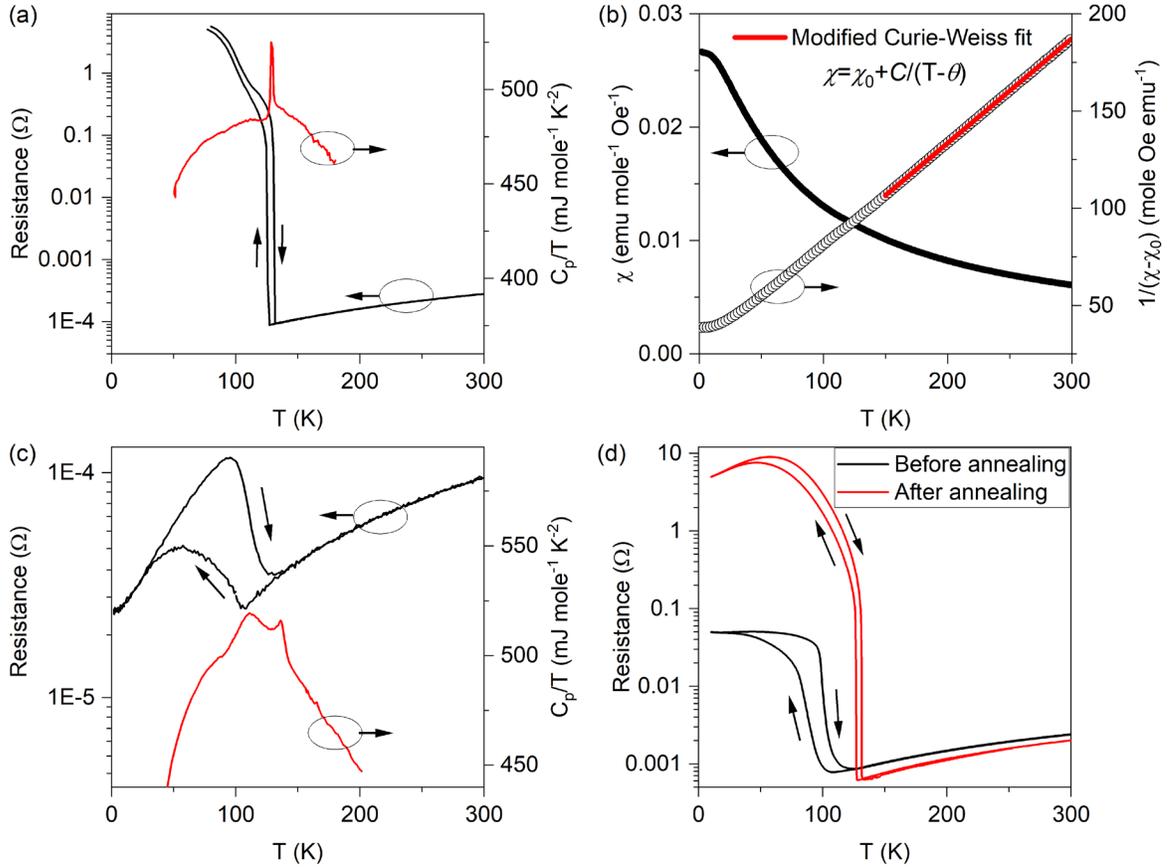

Figure 4. (a) Transport measurement (left) and specific heat (right) for sample A. (b) DC susceptibility χ (left) vs temperature and inverse magnetic susceptibility $1/(\chi-\chi_0)$ (right) vs temperature for sample A. (c) Transport measurement (right) and specific heat (left) for sample B. (d) Transport measurements of as-grown and high pressure post annealing for sample C.

**Physical properties.** We explored the physical properties of several specimens extracted from the boule. Among the samples studied, we found two classes of temperature-dependent transport behavior in the crystals. Data are presented here for three such samples, labeled A, B, and C. Electrical transport and specific heat data are plotted in Figure 4a for sample A. Because of the irregularity in the shape of the samples, we plot resistance rather than resistivity. Sample A exhibits a sharp metal-insulator transition (MIT) at $T_{MI}$~128 K spanning more than three orders of magnitude. This behavior is consistent with literature data reported on single crystals [38], although there the dynamic range is smaller (~$10^2$). Correspondingly, the specific heat exhibits one sharp first-order-like peak at 128 K as shown in Figure 4a. The temperature dependence of both DC magnetic susceptibility ($\chi$) and the inverse magnetic susceptibility ($1/(\chi-\chi_0)$) are shown in Figure 4b. No anomaly, discontinuity, or change in slope, which would mark the onset of magnetic ordering, was observed down to 2 K as Saito et al. mentioned [38]. This is likely because the susceptibility is dominated by the paramagnetic signal from $Pr^{3+}$. A Curie Weiss fit in the range from 150 K to 300 K, $\chi=\chi_0+C/(T-\theta)$, where $\chi_0$ represents $T$-independent contributions, $C$ and $\theta$ are Curie and Weiss constants, yields $\theta=-48$ K and $C=1.89$ emu K mole$^{-1}$ Oe$^{-1}$, respectively. The derived effective moment is $\mu_{eff}=3.9$ $\mu_B$/f.u., which is close to the 3.58 $\mu_B$/f.u. expected for a free ion $Pr^{3+}$ [44]. Other crystals, as exemplified by Sample B in Figure 4c, exhibit a clear transition in the transport around the MIT and then a re-entrance to a metallic behavior ($dR/dT > 0$) at lower temperatures. The DC magnetic susceptibility for sample B agrees with that of sample A. Samples exemplified by B showed a large temperature hysteresis, the width of which exhibited sample to sample variation. Such re-entrant behavior has previously been reported in polycrystalline samples, induced by high pressure, and attributed to oxygen non-stoichiometry [45]. To test whether oxygen deficiency was the cause of this behavior, we annealed Sample C, which showed a hysteresis in resistance, at $pO_2=200$ bar at 600 °C, followed by a slow cooling 0.5 °C/min to room temperature under 200 bar pressure. The resistance before and after annealing treatment are plotted in Figure 4d, which shows that the annealing yields a resistance akin to that exhibited by Sample A: a sharp transition with narrow hysteresis. This response suggests that the observed re-entrant behavior was a consequence of oxygen vacancies in the as-grown sample and that post-growth annealing reduces the vacancy concentration.

It is reasonable to question whether samples of type A or of annealed type C are indeed stoichiometric, $PrNiO_{3.0}$. We do not have a direct measure of this oxygen stoichiometry, However, we note that the MIT temperature for sample A and for annealed sample C are identical, suggesting that they share a common stoichiometry. It strikes us as unlikely that these two samples, prepared in different ways, would coincidentally have the same oxygen stoichiometry unless this stoichiometry was that of a terminal line phase. Of course, the transport measurement represents a bulk average, and some inhomogeneity may be present in one or both samples. Should this be the case, then the transport would be dominated by a fraction of the sample with composition $PrNiO_{3.0}$.

**X-ray scattering.** Motivated by the substantial interest in the community regarding the structural transitions in nickelates, we considered whether single crystal X-ray diffraction could be used to determine the symmetry of the low temperature and high temperature phases. This is complicated by the twinning of the crystal which causes the observed diffraction pattern to appear as a superposition of the patterns of all possible domains for a given distortion. It is thus most useful to view the structure in pseudocubic coordinates ($a=b=c\sim3.9$ Å); the lower symmetries, which possess larger unit cells than the pseudocubic cell, then generate additional Bragg reflections that that possess half-integer pseudocubic indices. Details of the transformation of systematic absences for twinned crystals can be found in the supplementary information.

Figure 5 shows the observed (hk0), (hk1), and (hk½) planes, each at 100 K, 150 K, and 300 K for sample A. The points marked by circles are forbidden for both the *Pbnm* and *P2$_1$/n* space groups, the points marked by triangles are forbidden by *Pbnm* but allowed for *P2$_1$/n*, and all other points indexed by integer/half-integer indices are allowed for both *Pbnm* and *P2$_1$/n*. Points at ($n/2,n/2,0$), where n is an odd integer, are systematically forbidden for both *Pbnm* and *P2$_1$/n* (Figures 5a-c), though they are strongest at 100 K and significantly weaker at 300 K. In principle, this could mean that the symmetry is lower than *P2$_1$/n* in both phases; however, other potential sources of these violators include multiple scattering (also known as Renninger reflections) and the possibility of a dilute presence of vacancy ordered superstructures of oxygen anions. Neither of these would be expected to have a strong temperature dependence, however, so it is useful to consider the temperature dependence of one of the stronger violators, such as (-5/2,-5/2,0), which is shown as one dimensional cuts along *k* in Figure 6a. The amplitude of (-5/2,-5/2,0) is comparable at 150 K and 300 K, but it is enhanced at 100 K. One possible interpretation of this temperature dependence is that the constant amplitude in the high temperature phase (150 and 100 K) is not intrinsic to the crystal (e.g. from multiple scattering), but that the enhancement in the low temperature phase is intrinsic. We note that recently Gawryluk et al. performed an analysis of the distortion modes, which suggested that the symmetry of the low temperature phase may be less than *P2$_1$/n* [46]. We considered the effect of removing the n-glide, but because of the twinning the systematic absences are no different than for *P2$_1$/n*. Thus, removal of the n-glide would not explain the violators that we observed in the (hk0) plane.

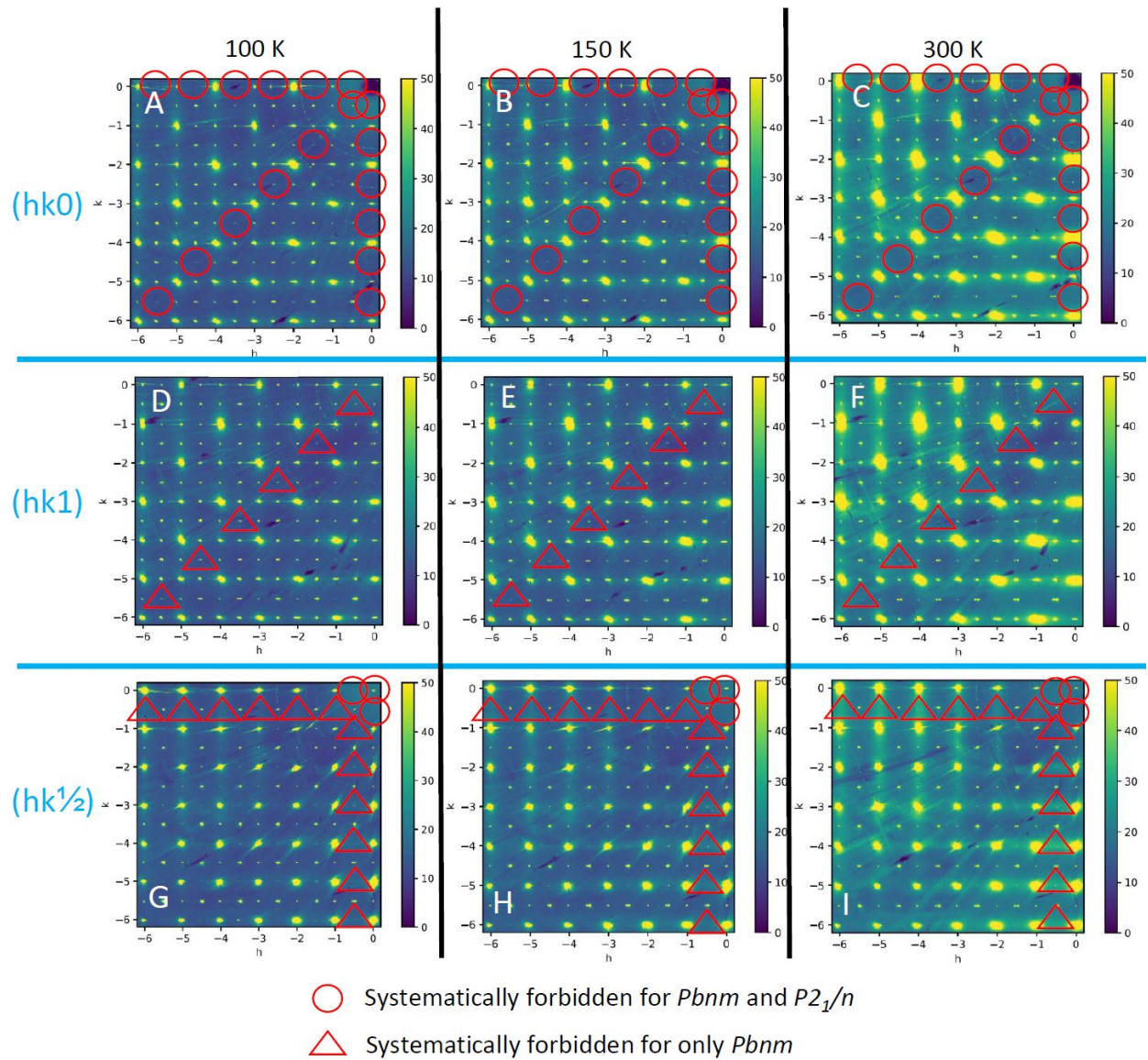

Figure 5. Pseudocubic (hk0), (hk1) and (hk½) planes measured by single crystal X-ray diffraction at 100 K, 150 K, and 300 K. Points that are forbidden by both *Pbnm* and *P2₁/n* space groups are circles, and points that are forbidden by *Pbnm* but allowed by *P2₁/n* are surrounded by triangles.

The (hk1) and (hk½) planes possess points that differentiate the *Pbnm* and *P2₁/n* space groups; thus the phase transition should generate some additional peaks in these planes. However, similar to the case of the (hk0) plane, the pattern of reflections appeared the same in both the low and high temperature phases, although enhancement of the peak amplitudes was observed in the low temperature phase. A particular example is the (-5/2,-5/2,1) point, which is shown as one-dimensional cuts along k in Figure 6b. Whereas the peak is weak at 300 K, it is somewhat stronger at 150 K, and significantly stronger at 100 K. Again, a potential interpretation is that the high temperature component is extrinsic, whereas the low temperature component is intrinsic, though if this interpretation is correct then the enhancement at 150 K, which should be well into the high temperature phase, is puzzling. Ultimately, we could not come to a firm conclusion regarding the intrinsic symmetries of the low and high temperature phases because of the combined effects of twinning and potential multiple scattering; diffraction from a single untwinned grain would be highly favorable for addressing the correct space group and structure.

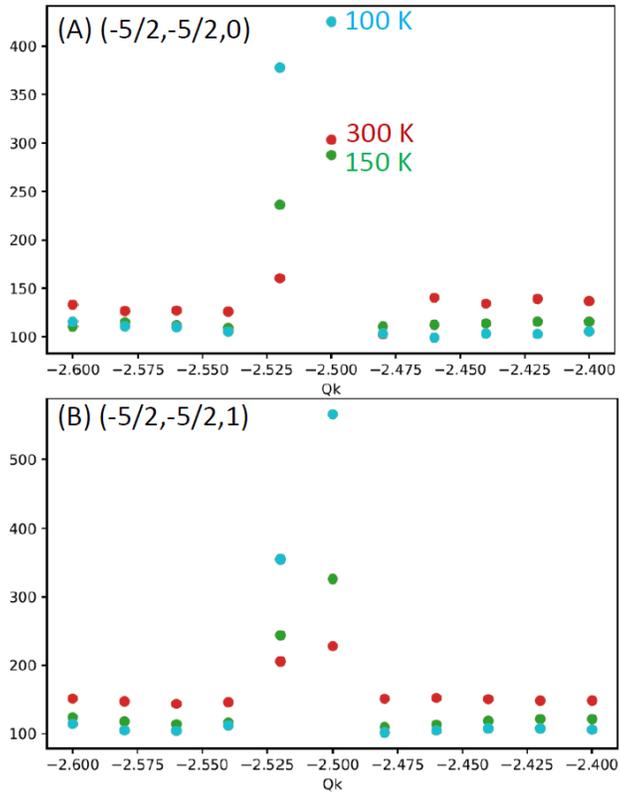

Figure 6. Cuts along k through the (-5/2,-5/2,0) and (-5/2,-5/2,1) points, shown in A and B respectively, at 100 K, 150 K, and 300 K. These cuts were made over integration ranges of 0.04 reciprocal lattice units in both h and l.

**Conclusion**

Single crystals of perovskite nickelate $PrNiO_{3-x}$ have been grown under an oxygen pressure of 295 bar using a unique high-pressure optical-image floating zone furnace. Single crystal and powder X-ray diffraction, resistivity, specific heat, and magnetic susceptibility measurements have confirmed the phase and quality. We found two types of behavior from the transport measurements, a MIT at $T_{MI}$~128 K and a clear transition in the transport around $T_{MI}$ and a re-entrance to a metallic state at lower temperatures. The re-entrant behavior can be modified by high pressure oxygen post – growth annealing, which we attributed to filling of oxygen vacancies in the as-grown sample. Single crystal X-ray scattering evidenced extremely weak peaks that violated the systematic absence condition from *Pbnm* and *P2$_1$/n* and others that violated only *Pbnm*; these peaks were enhanced in the low temperature phase. However, it is unclear whether or not they are related to an actual breaking of the symmetry of stoichiometric PrNiO$_3$, and more work is needed on this front.

**Acknowledgement.** This work was supported by the US Department of Energy, Office of Science, Basic Energy Sciences, Materials Science and Engineering Division. This research used resources of the Advanced Photon Source, a U.S. Department of Energy (DOE) Office of Science User Facility operated for the DOE Office of Science by Argonne National Laboratory under Contract No. DE-AC02-06CH11357. This research has been supported in part by ORNL Postdoctoral Development Fund by UT-Battelle, LLC under Contract No. DE-AC05-00OR22725 with the U.S. Department of Energy.

# Supplementary Information for "High pO$_2$ Floating Zone Single Crystal Growth of the Perovskite Nickelate PrNiO$_{3-x}$"

Hong Zheng[1], Junjie Zhang[2], Bixia Wang[1], Daniel Phelan[1], M. J. Krogstad[1], Yang Ren[3], W. Adam Phelan[4], Omar Chmaissem[5], Bisham Poudel[5], J. F. Mitchell[1]

[1]Materials Science Division, Argonne National Laboratory, Argonne, Illinois 60439, USA

[2]Materials Science and Technology Division, Oak Ridge National laboratory, TN 37831, USA

[3]Advanced Photon Source, Argonne National Laboratory, Argonne, Illinois 60439, USA

[4]Platform for the Accelerated Realization, Analysis and Discovery of Interface Materials (PARADIM), Department of chemistry, The Johns Hopkins University, Baltimore, MD 21218, USA

[5]Department of Physics, Northern Illinois University, Dekalb, Illinois 60115

In the main text, we considered the single crystal X-ray diffraction patterns (Fig. 5) in pseudocubic symmetry. In doing so, we considered twinned orthorhombic (space group *Pbnm*) and monoclinic (*P2$_1$/n*) structures. When twinned, there are six matrix transformations that relate these two space groups to the pseudocubic axes:

$$M = \begin{bmatrix} -1/2 & 0 & 1/2 \\ 1/2 & 0 & 1/2 \\ 0 & 1/2 & 0 \end{bmatrix}, \begin{bmatrix} 1/2 & 0 & 1/2 \\ 1/2 & 0 & -1/2 \\ 0 & 1/2 & 0 \end{bmatrix}, \begin{bmatrix} 0 & 1/2 & -1/2 \\ 0 & 1/2 & 1/2 \\ 1/2 & 0 & 0 \end{bmatrix},$$

$$\begin{bmatrix} 0 & 1/2 & 1/2 \\ 0 & -1/2 & 1/2 \\ 1/2 & 0 & 0 \end{bmatrix}, \begin{bmatrix} 1/2 & -1/2 & 0 \\ 1/2 & 1/2 & 0 \\ 0 & 0 & 1/2 \end{bmatrix}, \begin{bmatrix} 1/2 & 1/2 & 0 \\ -1/2 & 1/2 & 0 \\ 0 & 0 & 1/2 \end{bmatrix},$$

such that $(\vec{a}, \vec{b}, \vec{c})_{pc} = (\vec{a}, \vec{b}, \vec{c})_{m,o} M$. The Miller indices transform by the same matrices such that $(h, k, l)_{pc} = (h, k, l)_{m,o} M$. The following reflection conditions exist for the Miller indices in the *Pbnm* space group: for 0*kl*, *k*=2*n* (where *n* is an integer); for *h*0*l*, *h*+*l*=2*n*; for *h*00, *h*=2*n*; for 0*k*0, *k*=2*n*; for 00*l*, *l*=2*n*. For the *P2$_1$/n* space group, the 0*kl* condition is deleted, and all other conditions remain. When calculating the positions of systematic absences that are marked in Fig. 5 of the main text, we transformed all allowed reflections for the *Pbnm* and *P2$_1$/n* space groups to the pseudocubic coordinates using the above matrices.